\documentclass[twocolumn,showpacs,floats,floatfix,superscriptaddress,aps,pra]{revtex4}
\usepackage{amsfonts}
\usepackage{amssymb}
\usepackage{amsmath}
\usepackage{graphicx}
\usepackage{bm}
\usepackage{color}
\usepackage{epsfig}
\usepackage{calc}
\usepackage{ifthen}
\usepackage{subfigure}
\usepackage{graphicx}
\usepackage{epstopdf}
\usepackage{epsfig}

\begin{document}

\title{Universal coupled theory for metamaterial Bound states in the continuum}
\date{\today }

\begin{abstract}
In this paper, we present a novel universal coupled theory for metamaterial Bound states in the continuum (BIC) or quasi-Bound states in the continuum (quasi-BIC) which provides ultra-high Q resonance for metamaterial devices. Our theory analytically calculates the coupling of two bright modes with phase. Our method has much more accuracy for ultra-strong coupling comparing with the previous theory (the coupling of one bright mode and one dark mode). Therefore, our theory is much more suitable for BIC or quasi-BIC and we can accurately predict the transmission spectrum of metamaterial BIC or quasi-BIC for the first time.
\end{abstract}

\pacs{42.82.Et, 42.81.Qb, 42.79.Gn, 32.80.Xx}
\author{Wei Huang}
\affiliation{Guangxi Key Laboratory of Optoelectronic Information Processing, Guilin University of Electronic Technology, Guilin 541004, China}

\author{Songyi Liu}
\affiliation{Guangxi Key Laboratory of Optoelectronic Information Processing, Guilin University of Electronic Technology, Guilin 541004, China}

\author{Jiaguang Han}
\affiliation{Center for Terahertz Waves and College of Precision Instrument and Optoelectronics Engineering, Tianjin University, Tianjin 3000072, China}

\author{Yu Cheng}
\affiliation{Guangxi Key Laboratory of Optoelectronic Information Processing, Guilin University of Electronic Technology, Guilin 541004, China}

\author{Shan Yin}
\email{syin@guet.edu.cn}
\affiliation{Guangxi Key Laboratory of Optoelectronic Information Processing, Guilin University of Electronic Technology, Guilin 541004, China}

\author{Wentao Zhang}
\email{zhangwentao@guet.edu.cn}
\affiliation{Guangxi Key Laboratory of Optoelectronic Information Processing, Guilin University of Electronic Technology, Guilin 541004, China}

\maketitle


\section{Introduction}
Bound states in the continuum (BIC) is initially proposed in quantum mechanics \cite{Neumann1929}, which trapped or guided modes with their frequencies in the frequency intervals of radiation modes \cite{Hsu2016}. Most recently, BIC has already been introduced for optical systems for the high-resonant phenomenon, such as photonic crystals \cite{Gao2016, Gansch2016, Yang2016, Bulgakov2008, Paddon2000, Pacradouni2000}, plasmonic structures \cite{Azzam2018}, optical waveguide coupler \cite{Yu2019, Chen2019, Lee2020}, Bragg gratings \cite{Bykov2019, Gao2019} and metamaterial \cite{Koshelev2018, Miroshnichenko2010, Kupriianov2019, Abujetas2019, Cong2015, Liang2020}. There are many practical applications for photonic systems, such as lasers \cite{Kodigala2017, Ha2018}, sensors \cite{Liu2017, Romano2018}, high-sensitive medical devices \cite{Khanikaev2013, Tittl2018} and filters \cite{Foley2014}. 

The BIC phenomenon can vastly increase the Q-value resonance, especially for the metamaterial, due to the high loss of a single metamaterial structure. The BIC comes from the substantial coupling between the two adjacent single metamaterial structures. Two uniform metamaterial structures have the same frequency and phase of lossy electromagnetic waves, and the lossy waves of two metamaterials have strong interference between each other. Therefore, two lossy electromagnetic waves interfere destructively, which produces the infinite Q-value resonance in the theoretical. In the practical metamaterial, two lossy electromagnetic waves can not be the same due to fabrication. Thus, we can not fabricate infinite Q-value resonance for the metamaterial. However, it can still provide a very high Q value for the metamaterial device, and can widely be used in various applications. 

Due to this widely practical BIC metamaterial application, the theory to understand BIC metamaterial is increasingly essential. Currently, there are few different theories to discuss the BIC in the photonic system, such as topological theory \cite{Zhen2015, Zhen2014}, Fano resonance theory \cite{Limonov2017,Gallinet2011}, Temporal Coupled-Mode Theory (TCMT) \cite{Yu2019, Chen2019, Lee2020} and coupled theory \cite{Cong2015}. 
However, those theories have their shortcomings. The topological theory employs the topology to explain where the infinite Q-valve comes from, but it can not predict how much Q-valve is. Fano resonance theory can predict the Q-valve, whereas it requires a fitting number q (Fano asymmetry parameter) which is not a physical and fundamental parameter for the metamaterial. TCMT comes from the two coupled mode of waveguides system, and the parameters of TCMT is not essential parameters for the metamaterial. Thus, TCMT is more suitable for the optical system. 
Recently, a remarkable paper \cite{Cong2015} employs the fundamental coupled theory for metamaterial to predict the Q-valve. However, they use the same idea for electrical impedance tomography (EIT), which only considers coupling between one bright mode and one dark mode. 
It brings the big issue which only works for low Q-valve situations because when the two metamaterial structures are becoming identical, the Q-value is increasingly larger and the external EM field excites two metamaterial structures. Thus, two metamaterial structures become two bright modes and the previously coupled theory \cite{Cong2015} is not valid anymore. 
The coupled theory is very fundamental theory and it widely used in many systems \cite{Huang2014, Huang2019, Huang2020, Huang20202}.

\begin{figure*}[htbp]
	\centering
		\includegraphics[width=0.32\textwidth]{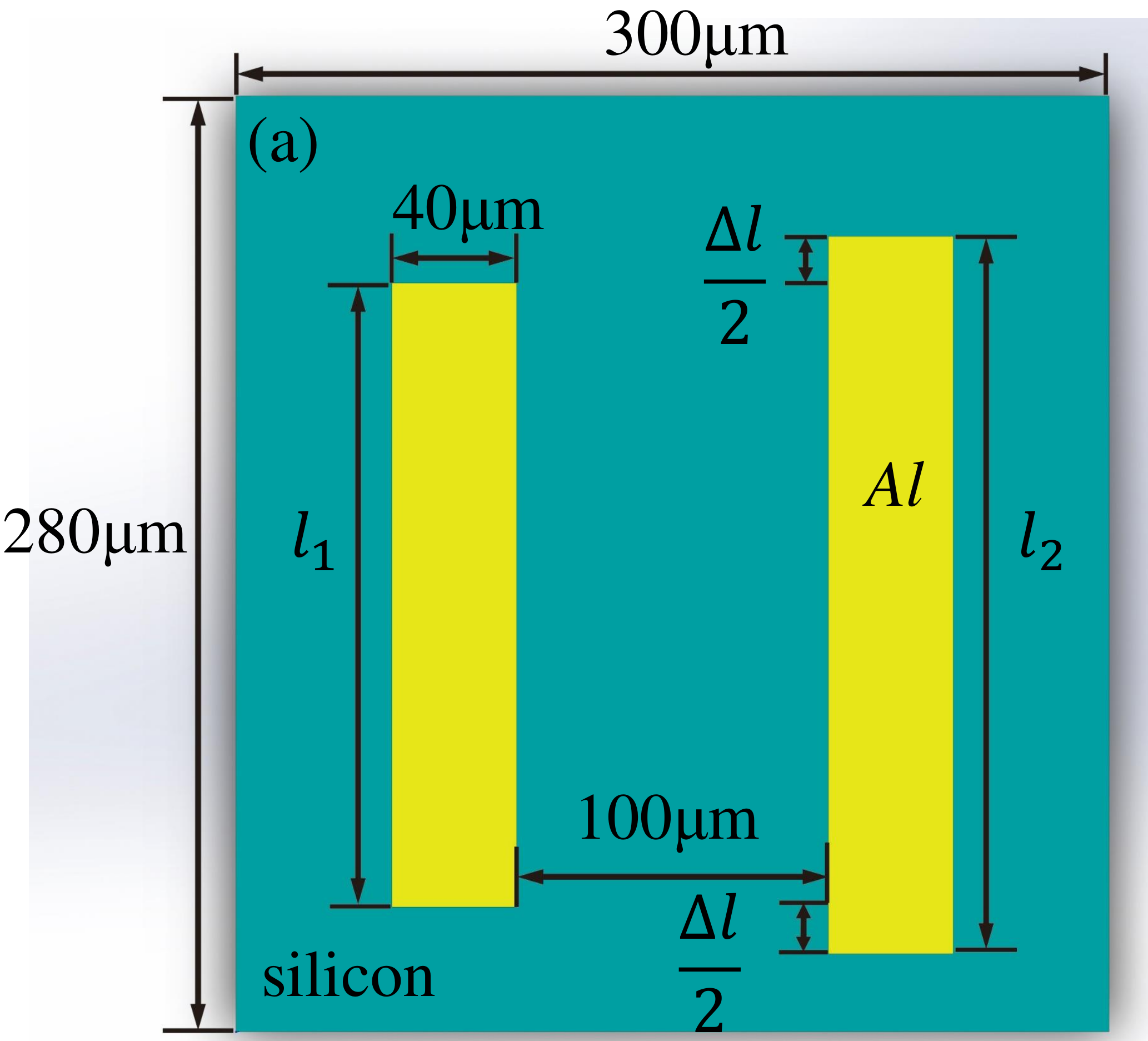}
		\includegraphics[width=0.325\textwidth]{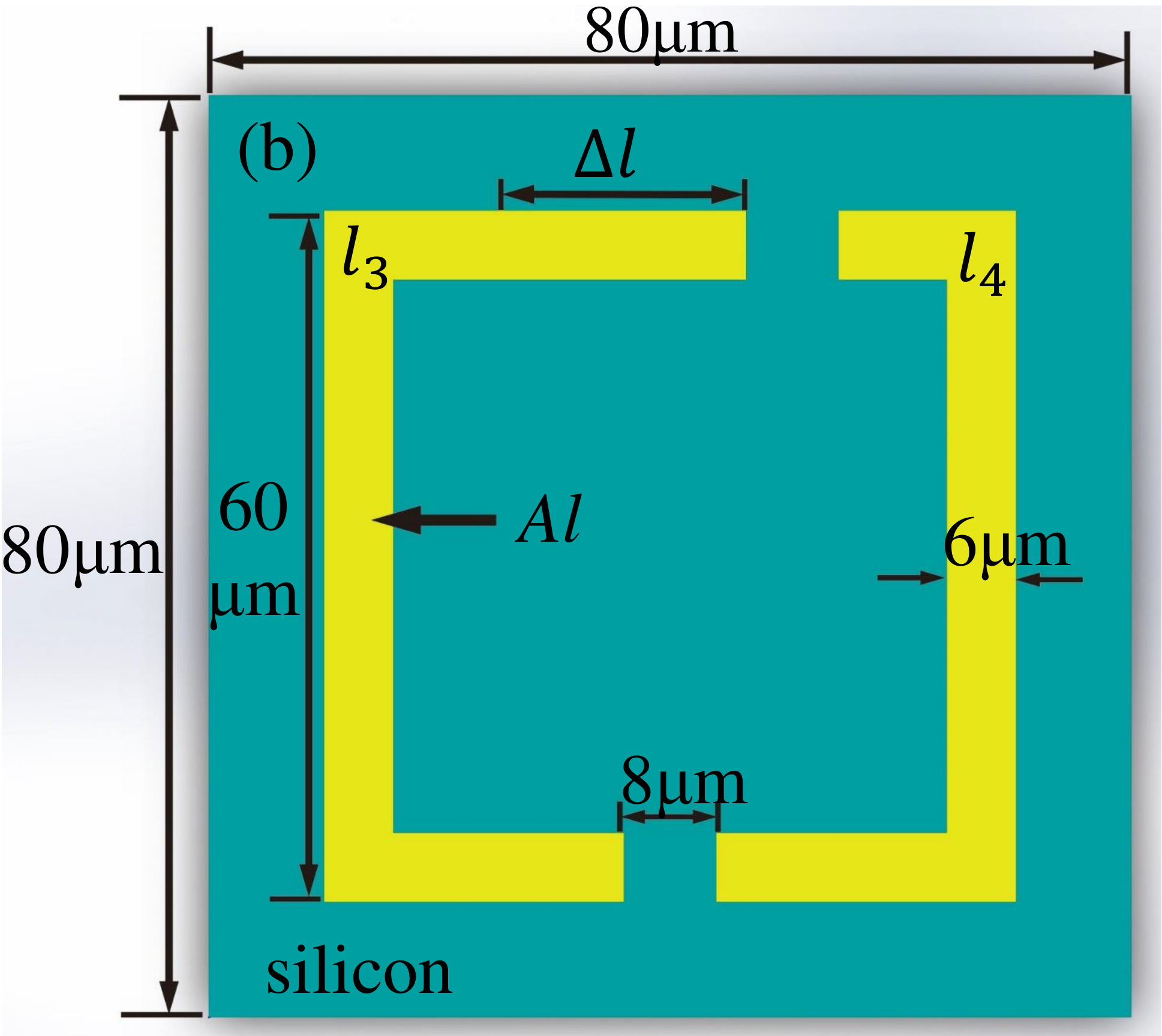}
		\includegraphics[width=0.32\textwidth]{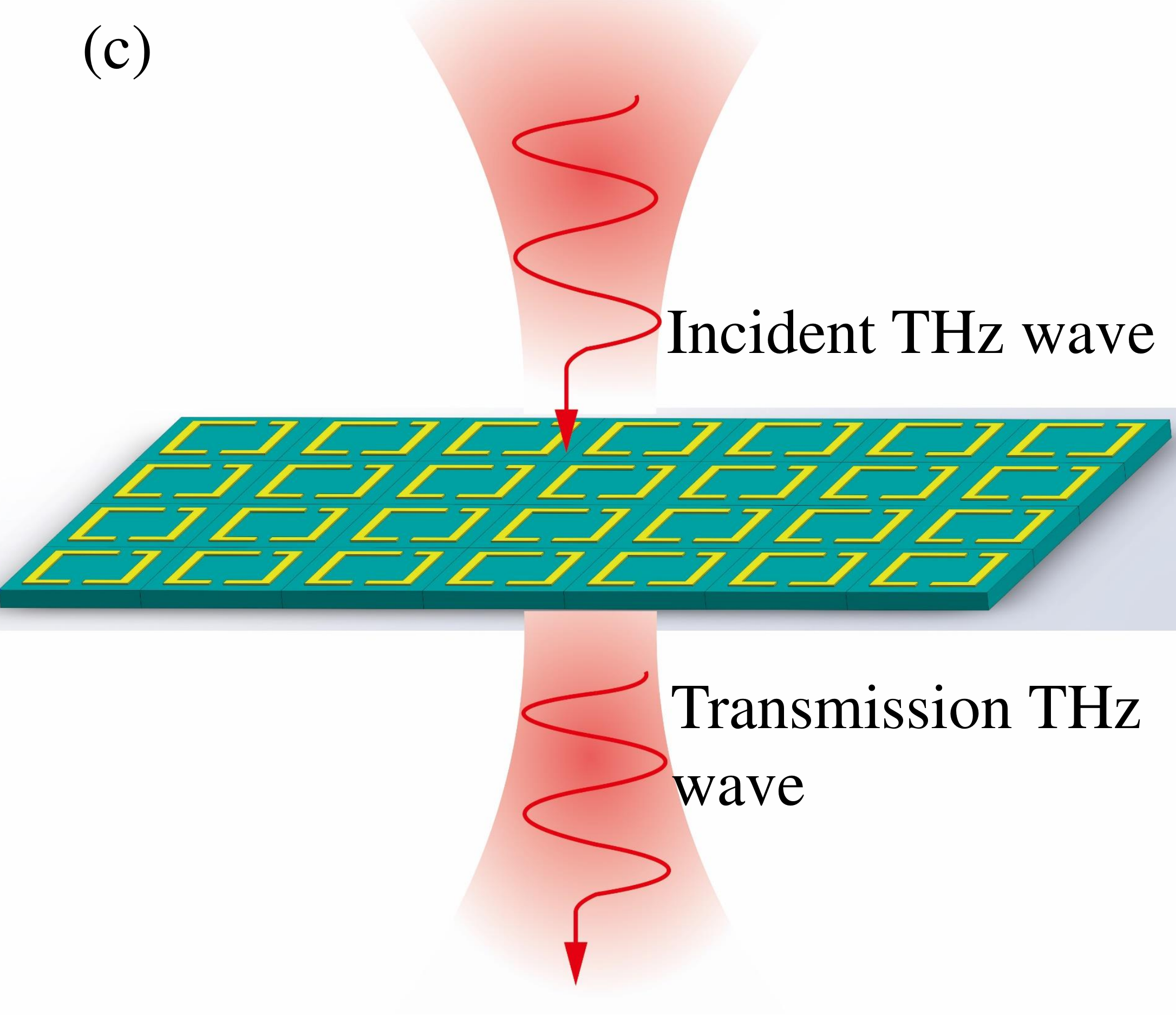}
	\caption{(a) The configurations of cut wires (CWs) BIC. (B) The configurations of split ring resonators (SRRs) BIC. (c) The schematic figures of input THz wave and transmission spectrum.}
	\label{Fig1}
\end{figure*}

This paper proposes a brand universal coupled theory for metamaterial BIC, which contains the coupling between two bright modes with phase. Therefore, our theory can predict a very high Q-valve situation and the parameters of our view come from the fundamental physical parameters of metamaterial, such as the resonance frequency $\omega_1$, $\omega_2$ for each single metamaterial structure, the loss $\gamma_1$, $\gamma_2$ for each single metamaterial structure and the phase $\phi_1$ and $\phi_2$ with each resonance frequency for each single metamaterial structure. The coupling strength $g$ comes from the physical configuration of two metamaterial structures. The beauty of our theory is that the resonance frequencies, the loss and the phases are obtained by the spectrum of each signal metamaterial structure, and coupling strength describes the connection of two metamaterial structures.

To demonstrate our theory, we employ two different types of BIC, such as coupling between two cut wires (CWs) \cite{zhang2014} (see Fig. \ref{Fig1} (a)) and two split ring resonators (SRRs) \cite{Cong2015} (see Fig. \ref{Fig1} (b)). We predict the frequency of BIC and Q-value from the spectrum of transmission THz wave, as shown in Fig. \ref{Fig1} (c). 

\begin{figure*}[htbp]
	\centering
		\includegraphics[width=0.4\textwidth]{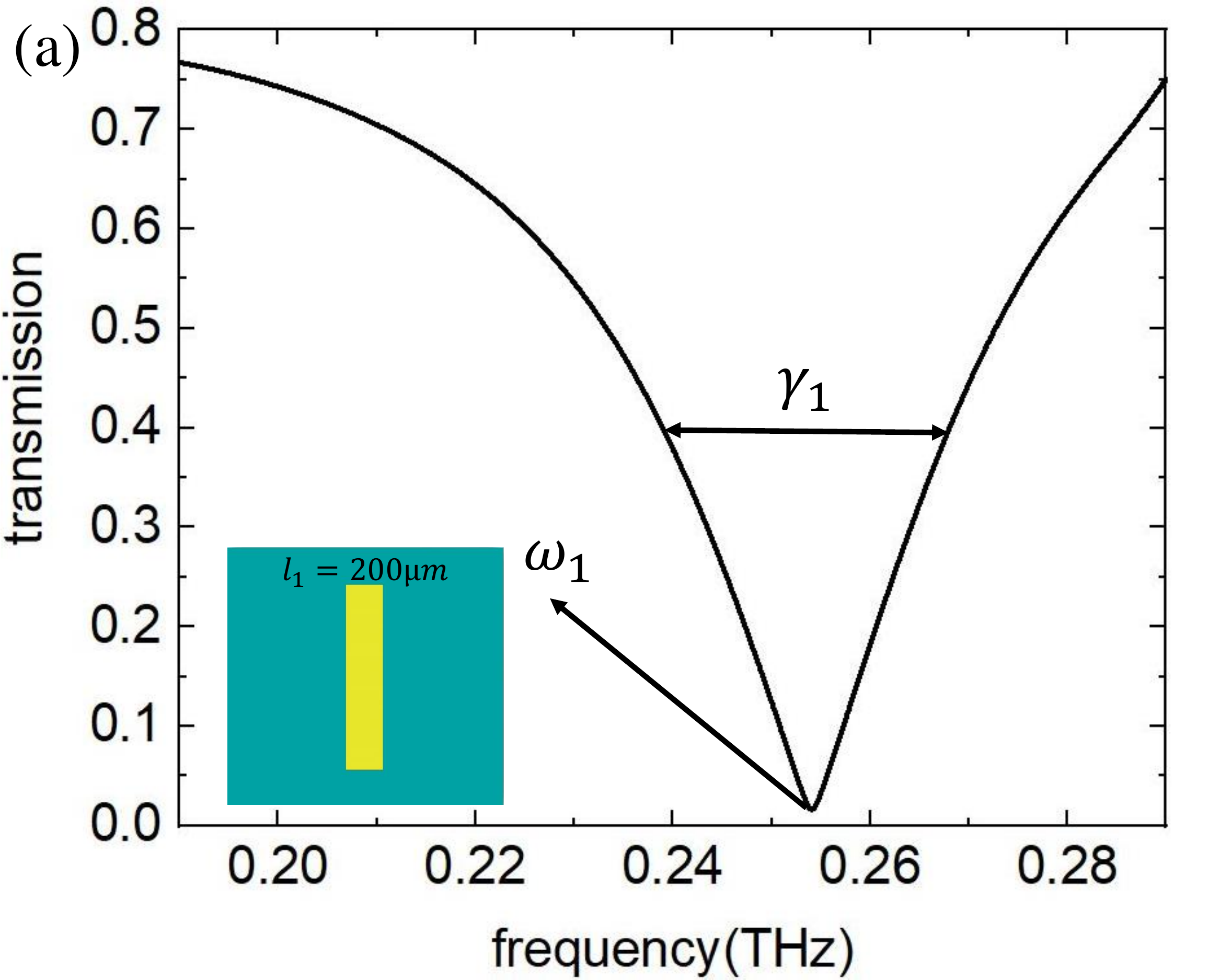}
		\includegraphics[width=0.405\textwidth]{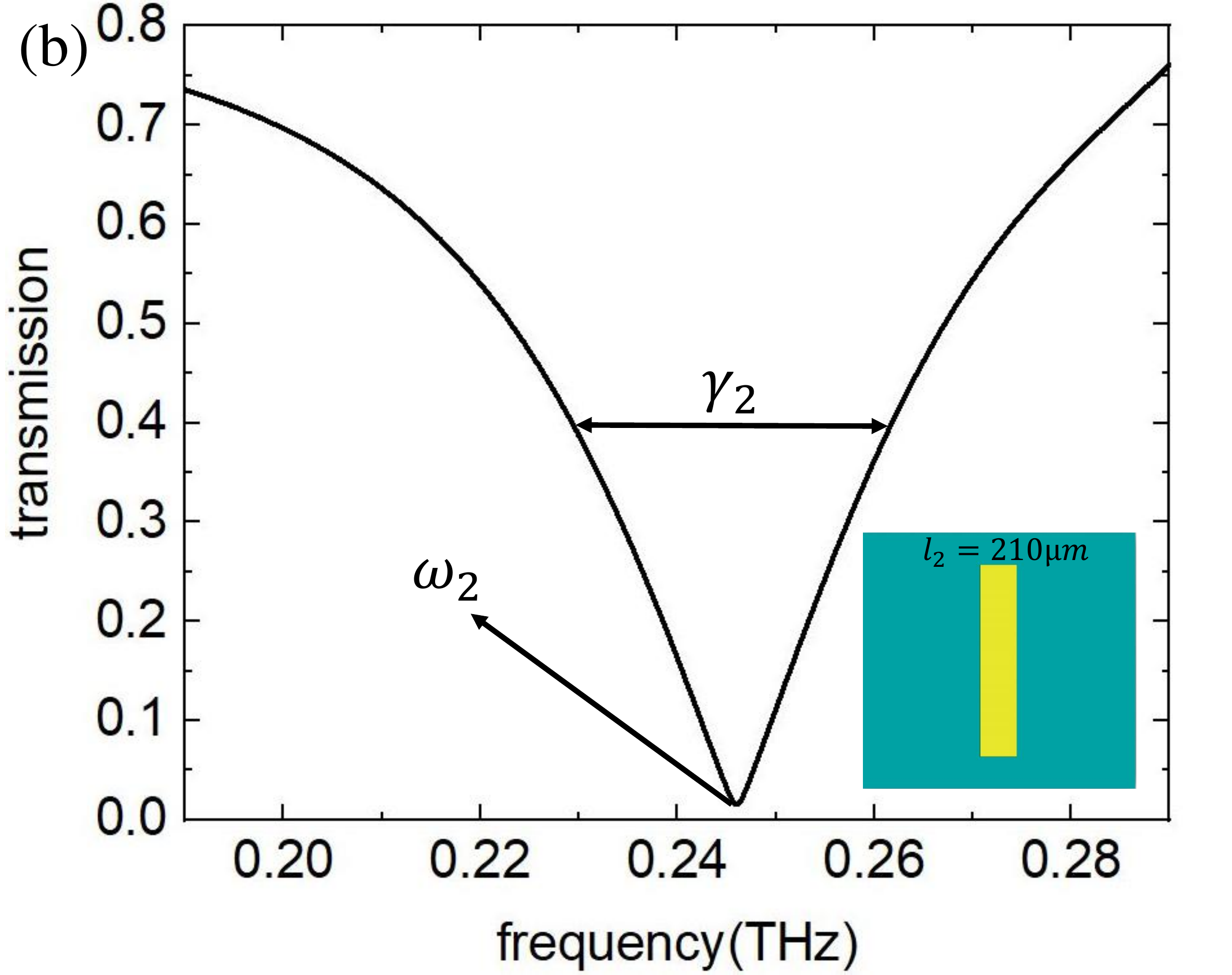}
		\includegraphics[width=0.4\textwidth]{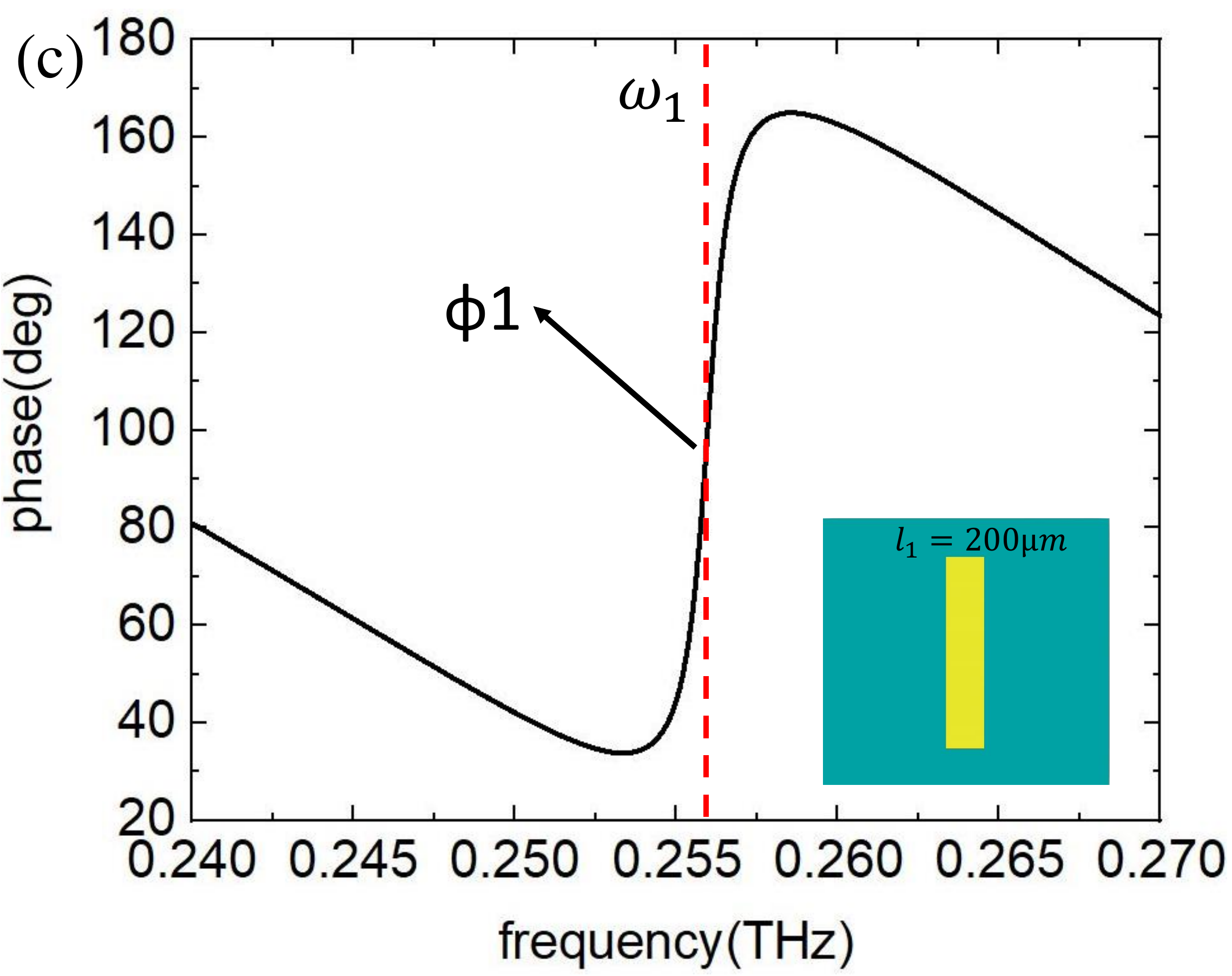}
		\includegraphics[width=0.4\textwidth]{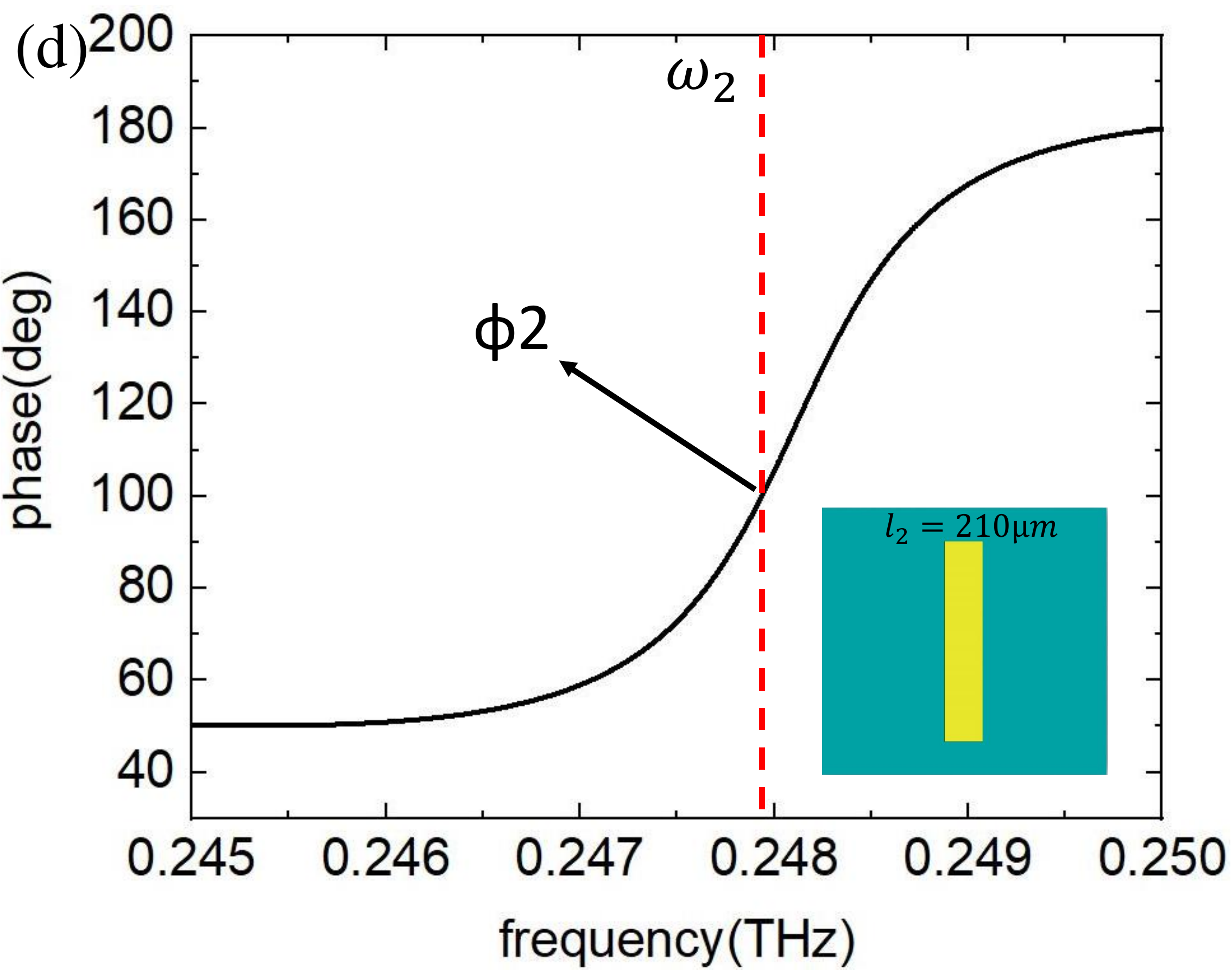}
	\caption{Full-wave simulations of (a) transmission spectrum of single left CW structure, (b) transmission spectrum of single right CW structure, (c) the phase of single left CW structure and (d) the phase of single right CW structure. }
	\label{Fig2}
\end{figure*}

\begin{figure}[htbp]
	\centering
		\includegraphics[width=0.45\textwidth]{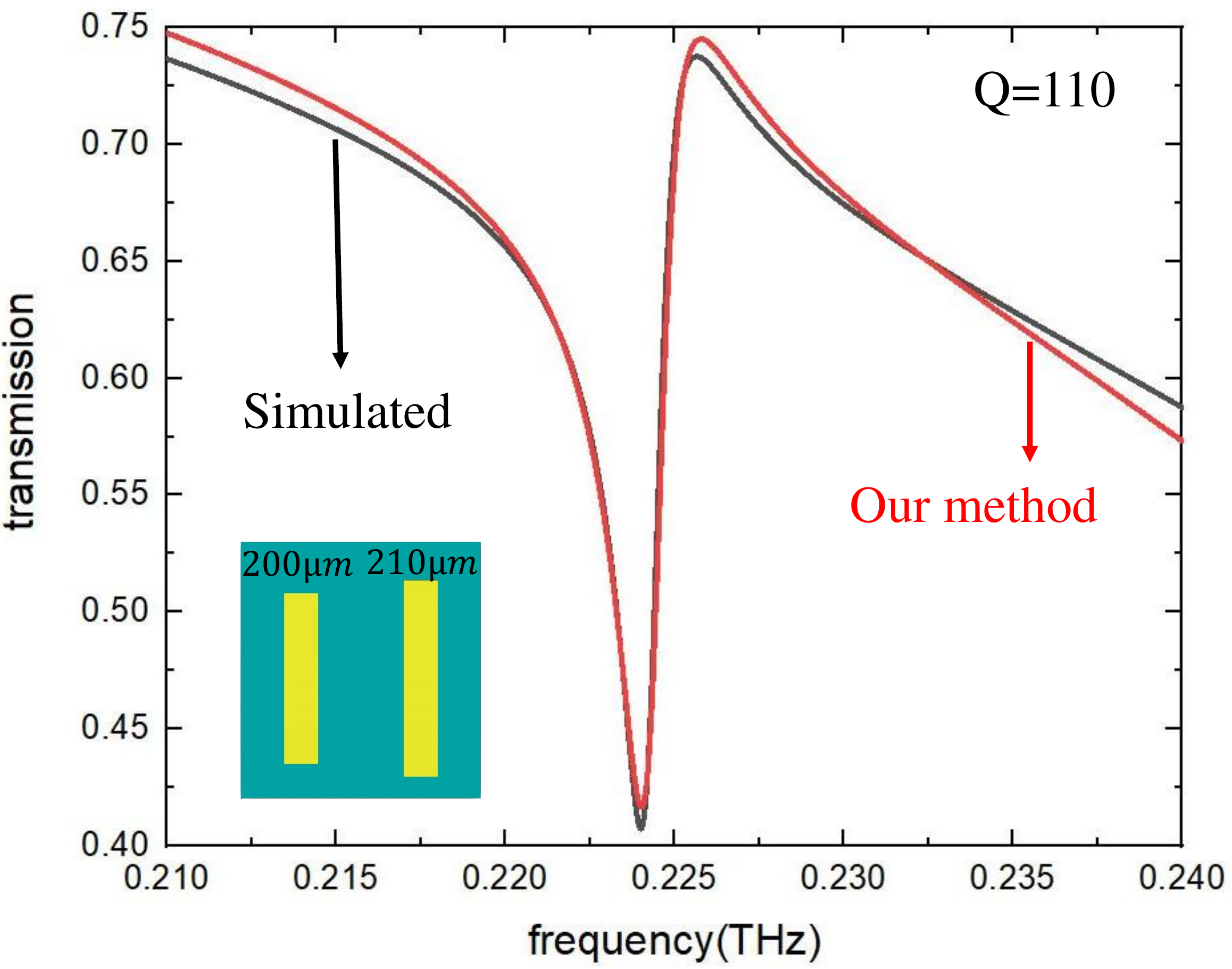}
	\caption{The black line is the full-wave simulations of cut wires (CWs) BIC (see the smaller figure) and our predict the transmission spectrum of cut wires (CWs) BIC by employing our theory, as shown in red line. }
	\label{Fig3}
\end{figure}

\begin{figure*}[htbp]
	\centering
		\includegraphics[width=0.4\textwidth]{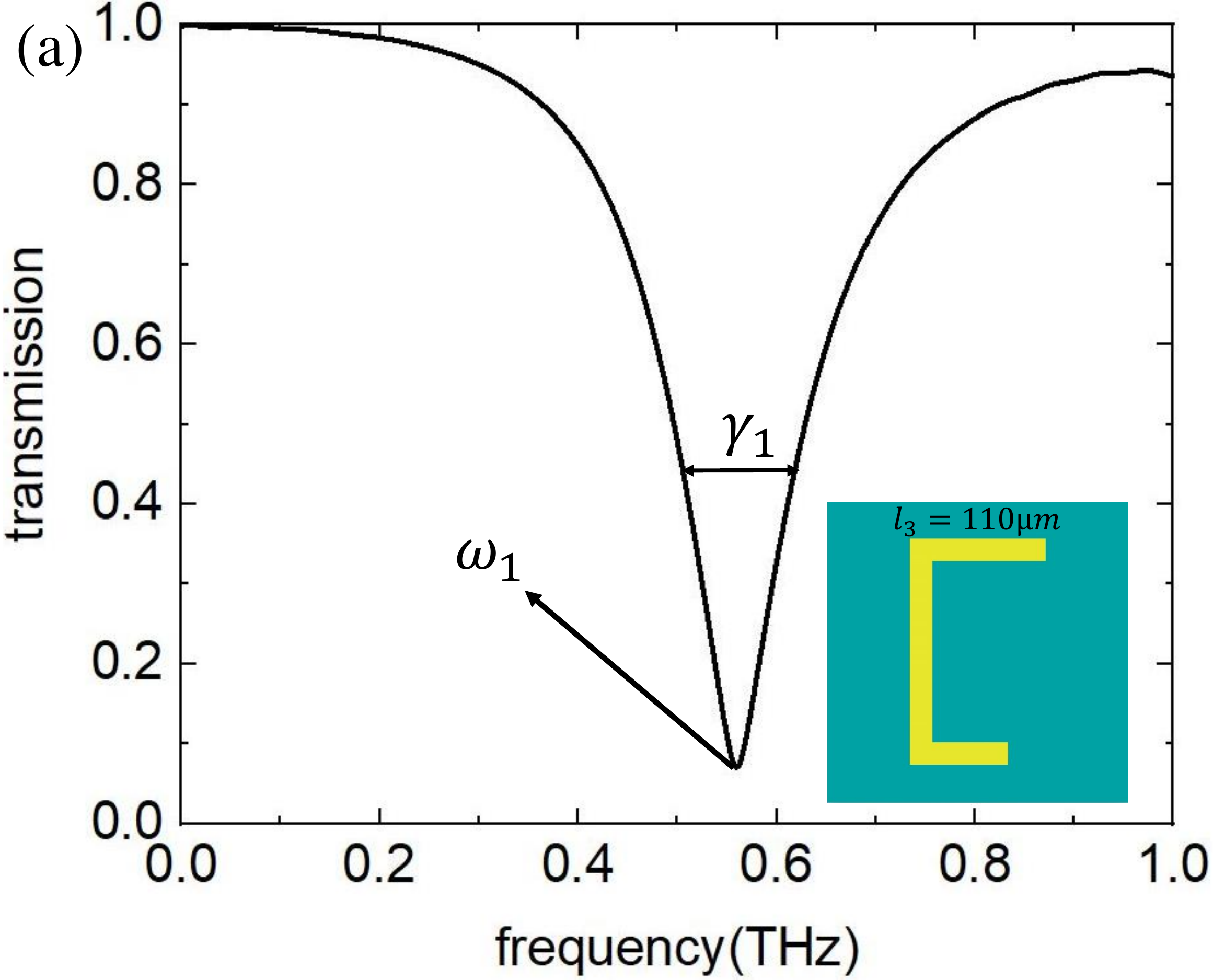}
		\includegraphics[width=0.4\textwidth]{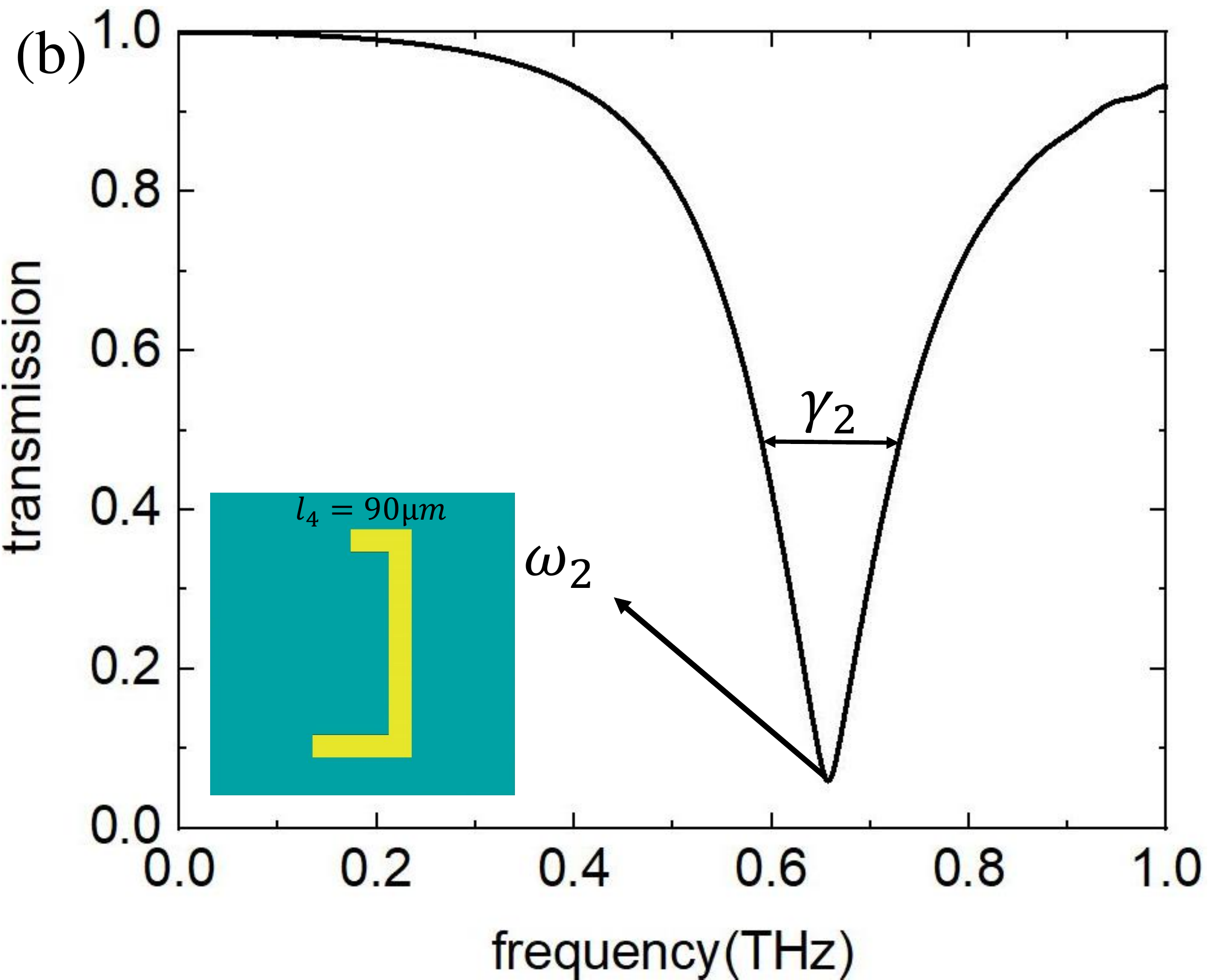}
		\includegraphics[width=0.4\textwidth]{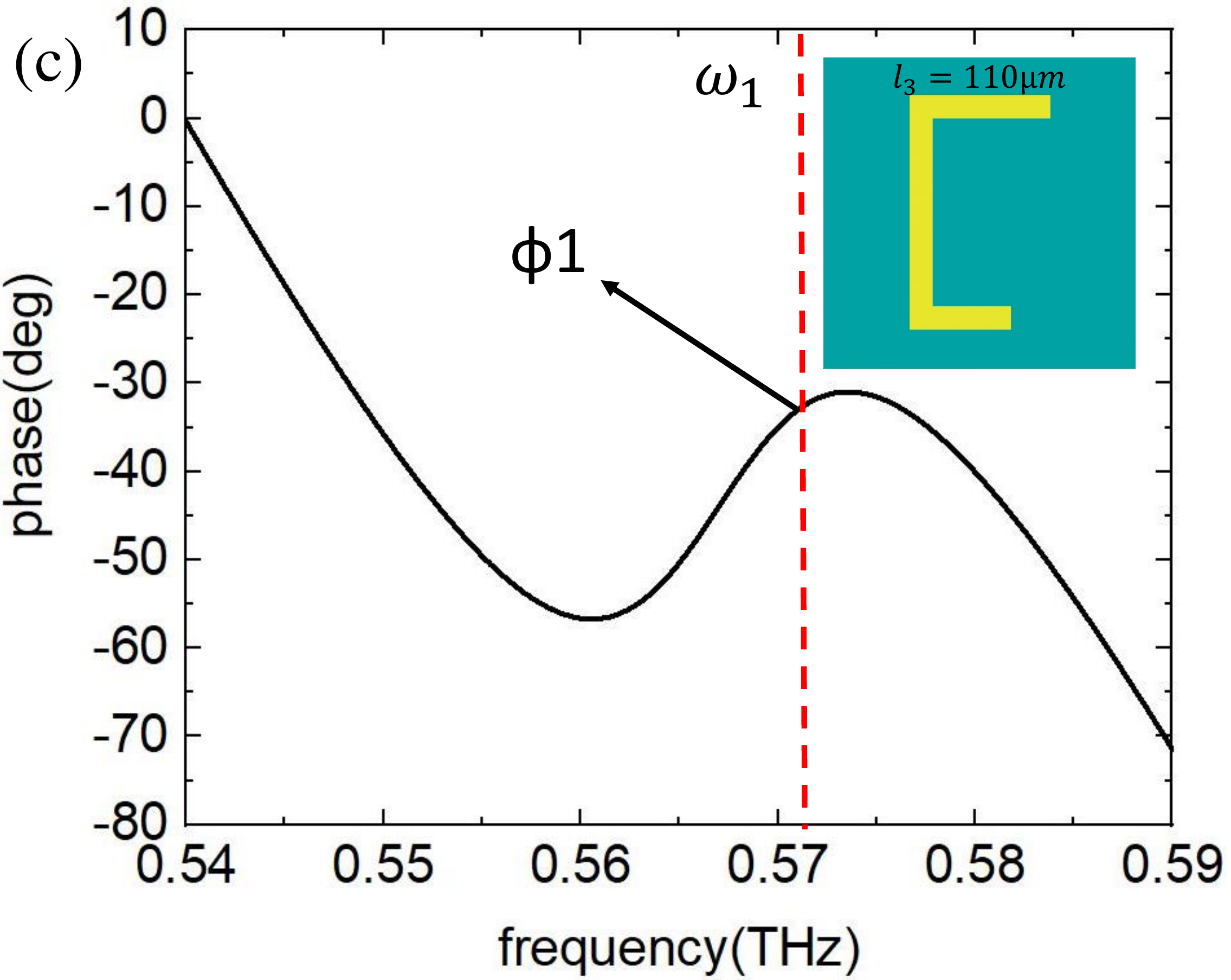}
		\includegraphics[width=0.4\textwidth]{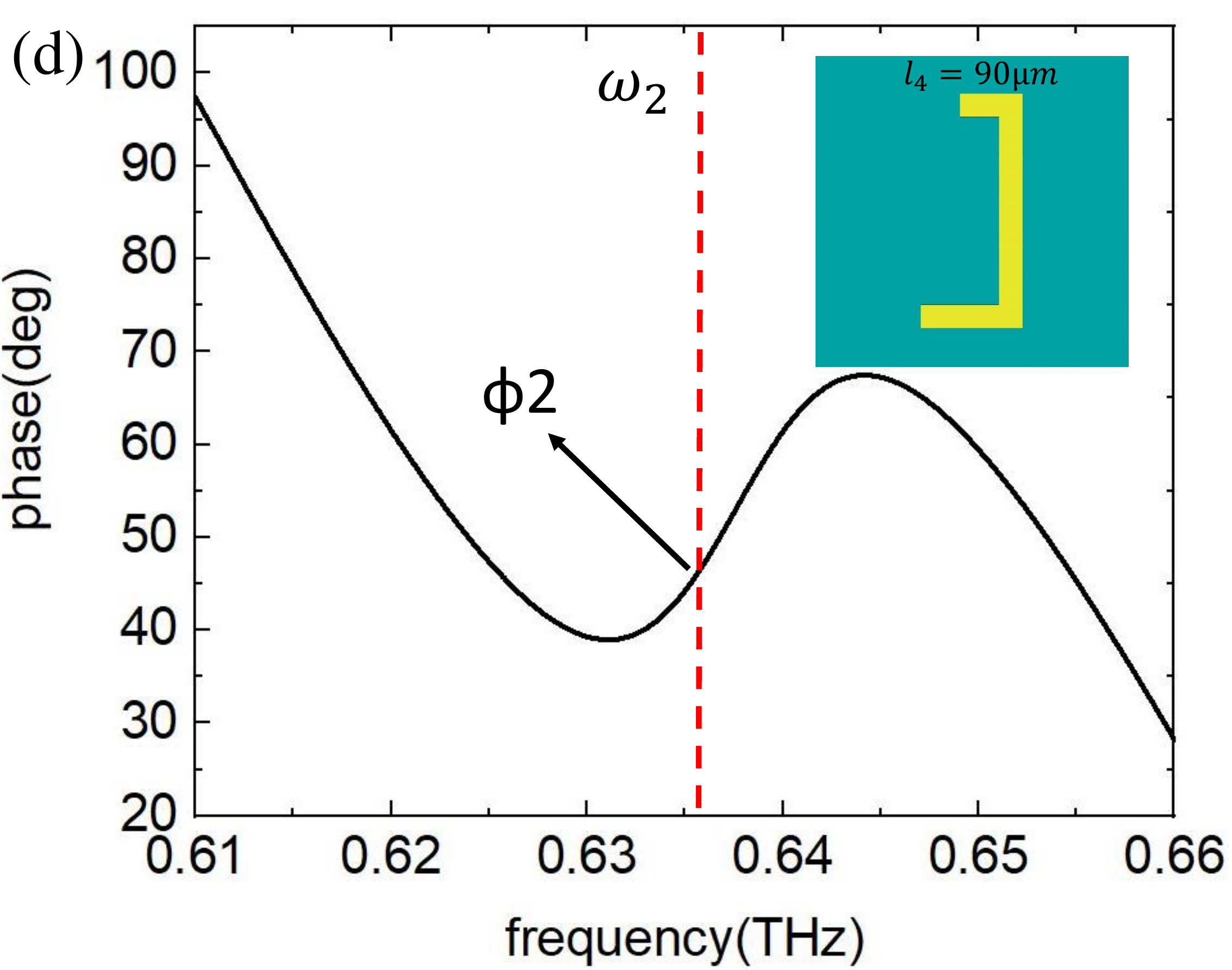}
	\caption{Full-wave simulations of (a) transmission spectrum of single left SRR structure, (b) transmission spectrum of single right SRR structure, (c) the phase of single left SRR structure and (d) the phase of single right SRR structure. }
	\label{Fig4}
\end{figure*}

\begin{figure}[htbp]
	\centering
		\includegraphics[width=0.45\textwidth]{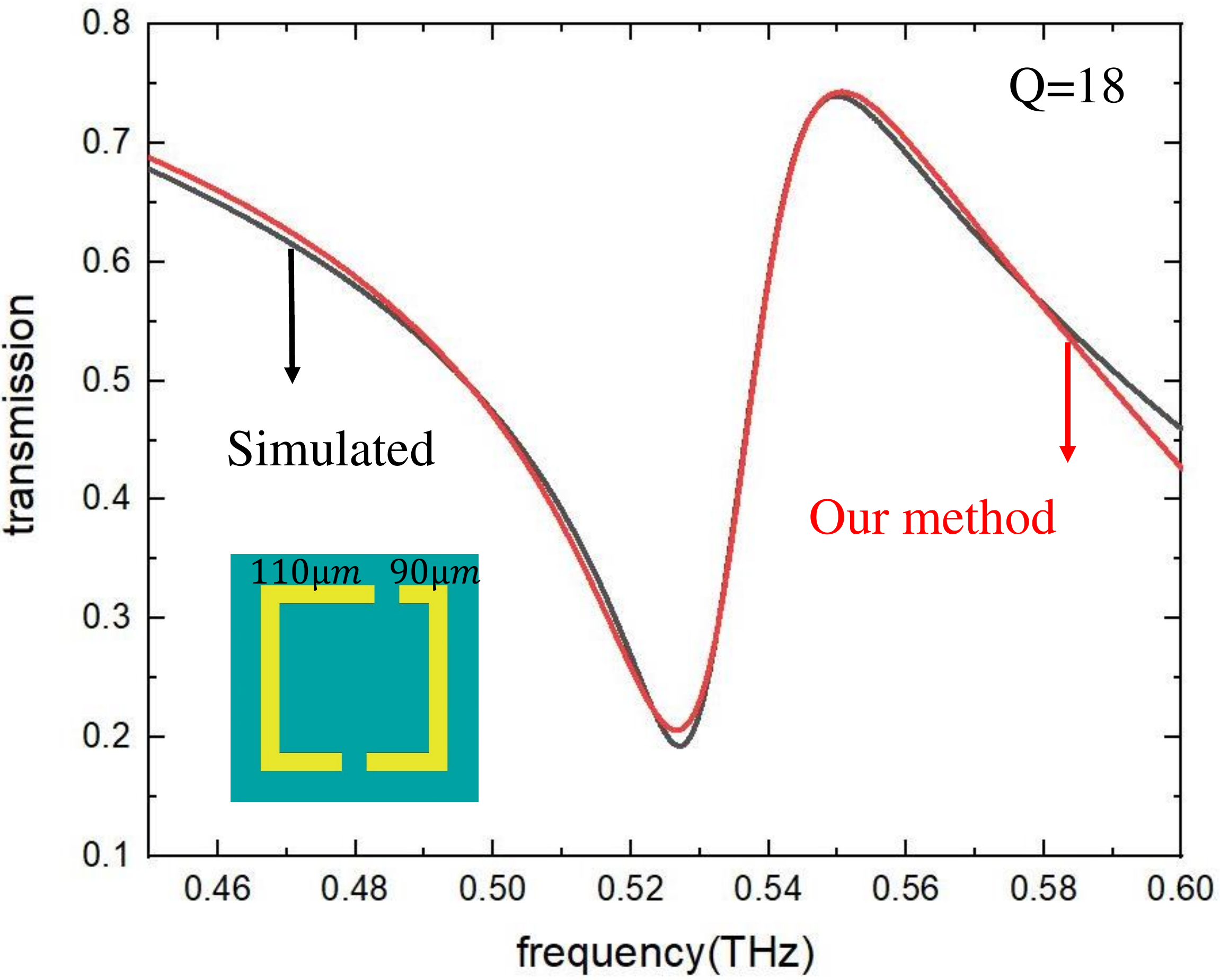}
	\caption{The black line is the full-wave simulations of split ring resonators (SRRs) BIC (see the smaller figure) and our predict the transmission spectrum of split ring resonators (SRRs) by employing our theory, as shown in red line. }
	\label{Fig5}
\end{figure}

\section{Universal coupled theory}

In this section, we present a brand \textit{universal coupled theory} to describe the BIC in the metamaterial. The idea of our theory comes from the spectrum of transmission THz wave, which is the linear superposition of the transmission spectrum of each metamaterial structure. The linear superposition has a very close relationship with energies in each metamaterial structure. We take the $|a|^2$ and $|b|^2$ as the energies in each metamaterial structure. Furthermore, the energies in each metamaterial structure can be well described by coupled theory. Besides, the phases of each metamaterial structure become increasingly important for strong coupling and BIC. Thus, we should introduce the phase information into the coupled theory. Finally, we obtain our \textit{universal coupled theory} as following, 

	\begin{equation}
\left[
	\begin{matrix}
	\ w-w _{a}-i\gamma_{a} & \Omega \\
	\Omega & \ w-w _{b}-i\gamma_{b}%
	\end{matrix}%
	\right] \left[
	\begin{matrix}
	a \\
	b%
	\end{matrix}%
	\right] =\left[
	 \begin{matrix}
	 \sqrt{\gamma_{a}}E \\
	 \sqrt{\gamma_{b}}e^{_{i\phi}}E%
	 \end{matrix}%
	 \right],
	\end{equation}%

where $\omega$ is the frequency of input THz wave. $\omega_a$ and $\omega_b$ represent the frequencies of the metamaterial structures, which is the same as resonant frequency of corresponding  metamaterial structure ($\omega_a$ = $\omega_1$; $\omega_b$ = $\omega_b$). $\Omega$ is the coupling strength with loss, due to the loss of transferring energy from one metamaterial structure to another, with $\Omega = g-i\sqrt{\gamma_{a} \gamma_{b}}e^{_{i\phi}}$, where $g$ is the coupling strength between two metamaterial structures. $\gamma_a$ and $\gamma_b$ are the loss terms of metamaterial structures, which is closely to $\gamma_1$, $\gamma_2$ for each single metamaterial structure ($\gamma_{a}=\gamma_{1}$;  $\gamma_{b}=(\gamma_{1} - \gamma_{2})/2$). $\phi$ is the phase information, which can be calculated by the phase $\phi_1$ and $\phi_2$ for each metamaterial structure ($\phi=(\phi_{1} - \phi_{2})d$, where $d$ is the width of metamaterial structure). $E$ is the amplitude of external exciting THz wave.

Subsequently, we can obtain the energy amplitudes $a$, $b$ of each metamaterial structure by solving the Eq. 1, as shown,
\begin{equation}
	a =\frac{ ((w-w _{b}-i\gamma_{b}) \sqrt{\gamma_{a}}-\Omega \sqrt{\gamma_{b}}e^{_{i\phi}})E} { ( w-w _{b}-i\gamma_{b}) (w-w _{a}-i\gamma_{a})-\Omega^{2}};
\end{equation}
	
\begin{equation}
	b =\frac{ ((w-w _{a}-i\gamma_{a}) \sqrt{\gamma_{b}}-\Omega \sqrt{\gamma_{a}})E} { ( w-w _{b}-i\gamma_{b}) (w-w _{a}-i\gamma_{a})-\Omega^{2}}.
\end{equation}

Then we calculate the effective susceptibility which is the linear superposition with energy amplitudes $|a|^2$ and $|b|^2$. the effective electric susceptibility of the metamaterial can be written as \cite{meng2012}

\begin{figure*}[htbp]
	\centering
		\includegraphics[width=0.32\textwidth]{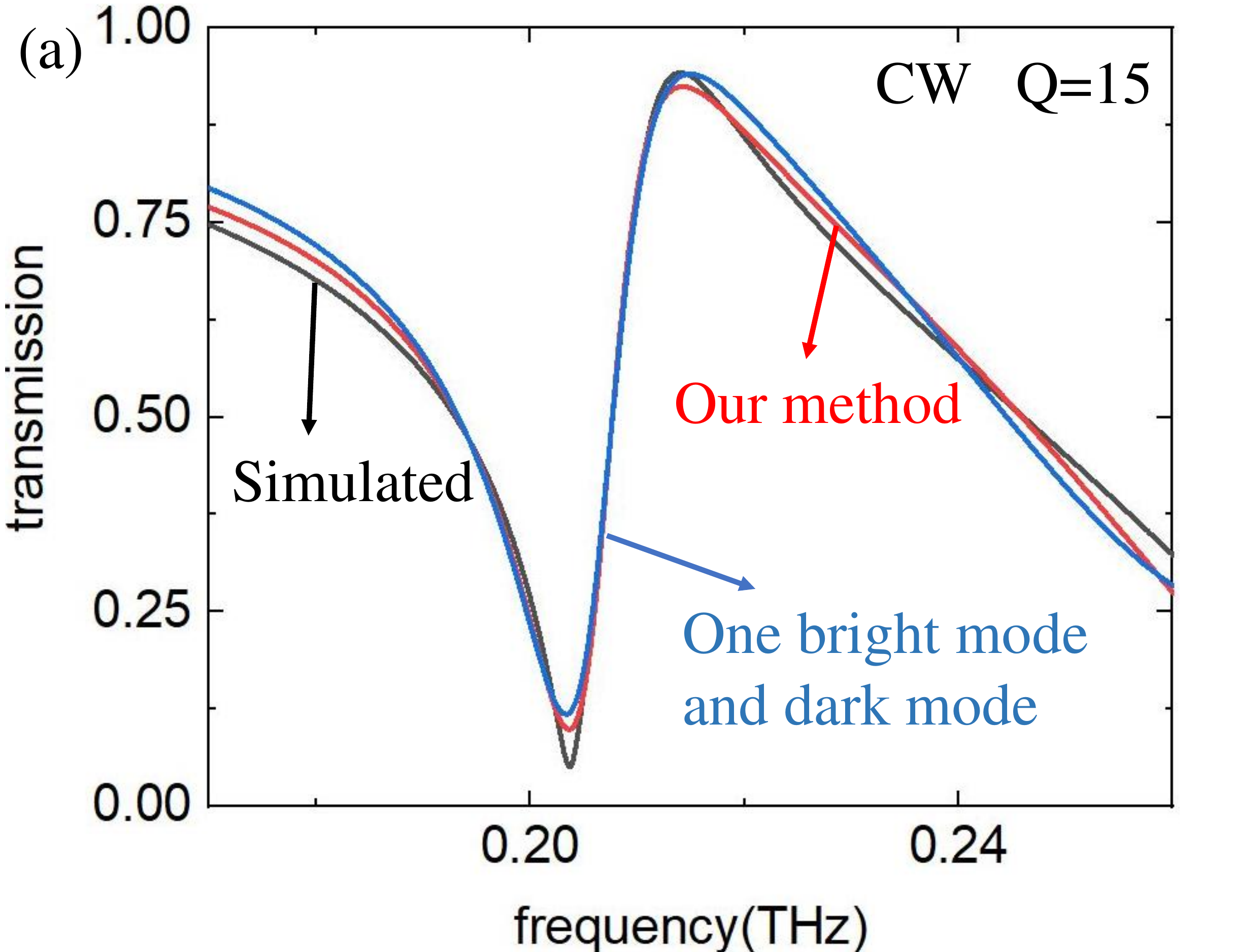} 
		\includegraphics[width=0.31\textwidth]{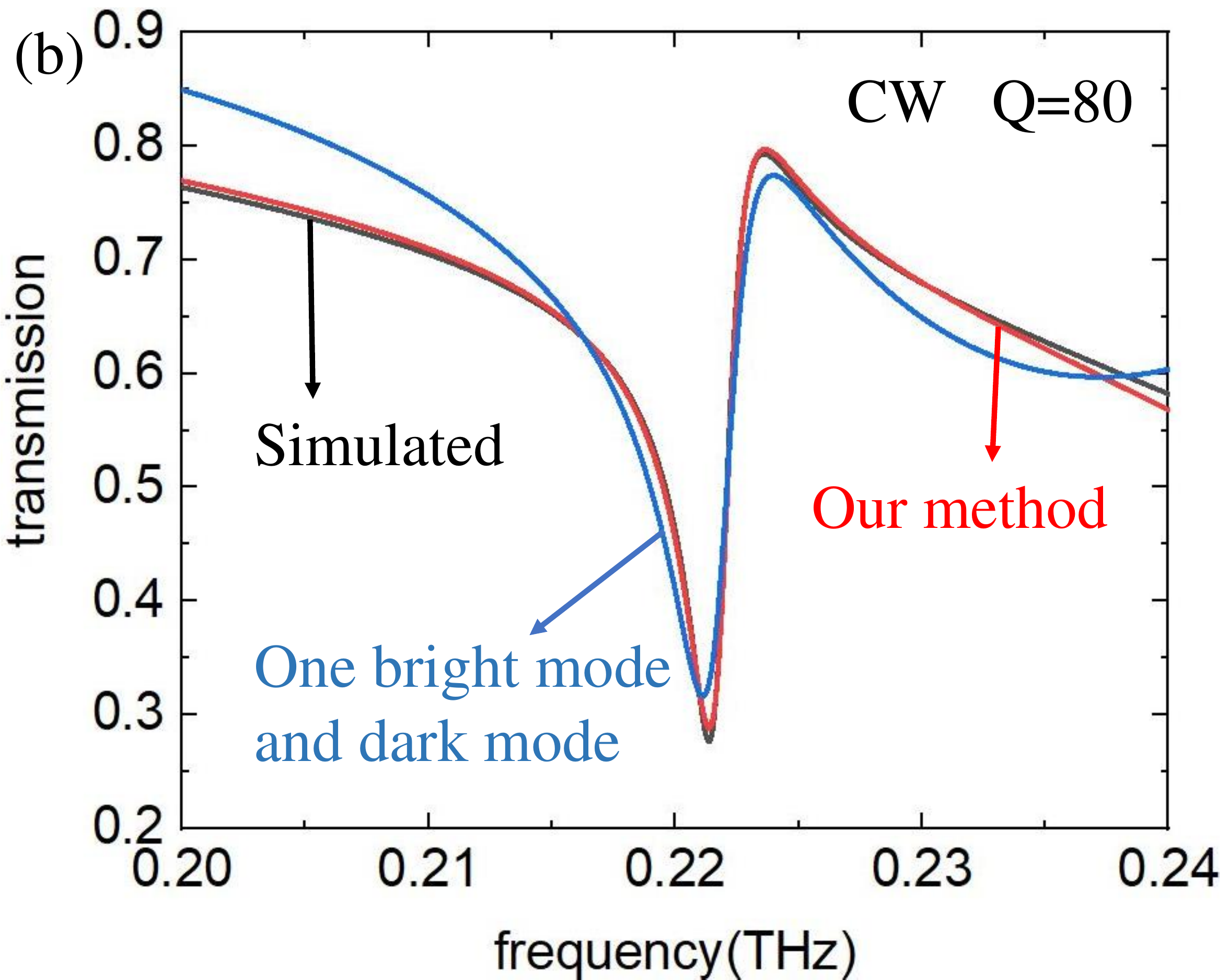} 
		\includegraphics[width=0.316\textwidth]{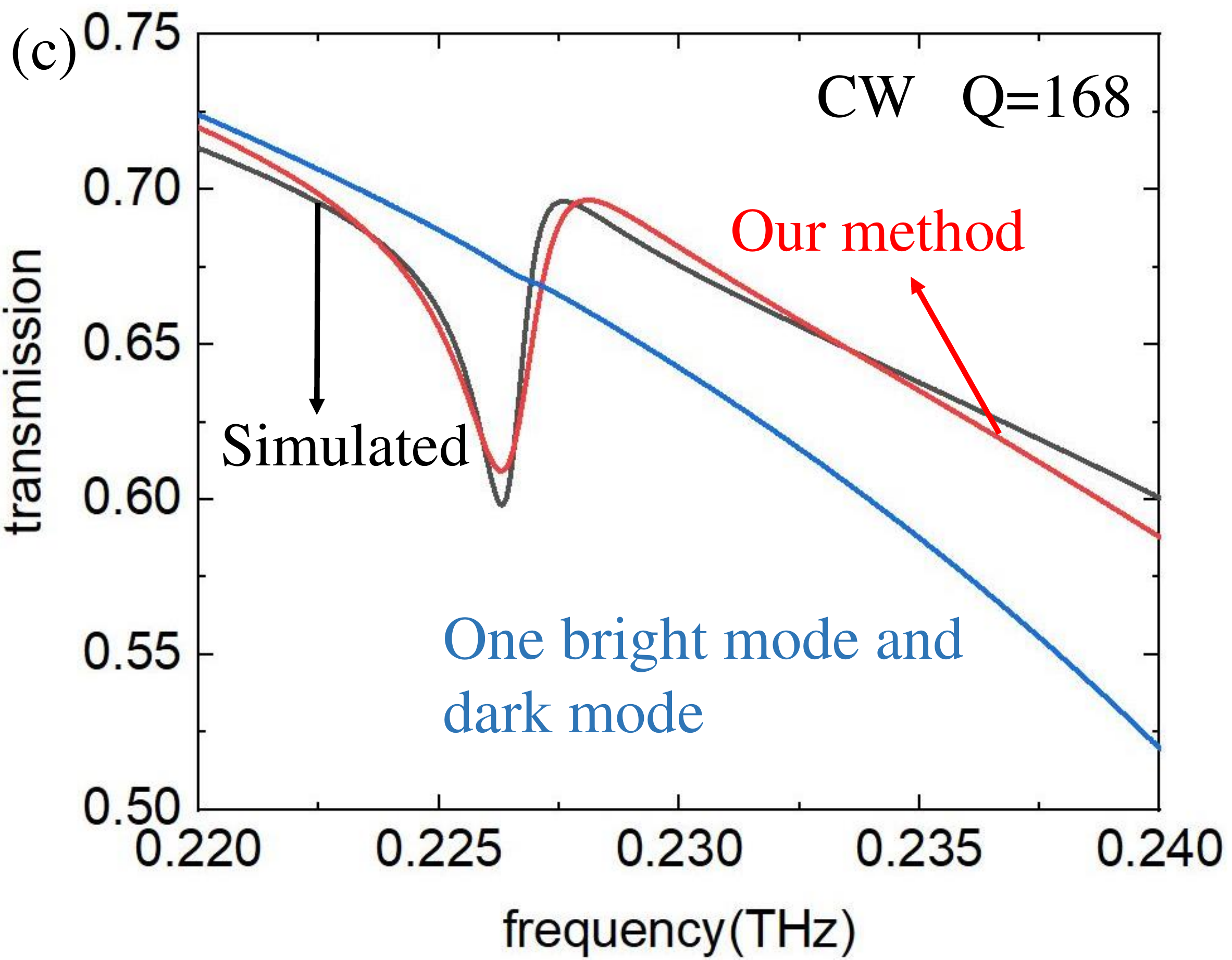}
		\includegraphics[width=0.32\textwidth]{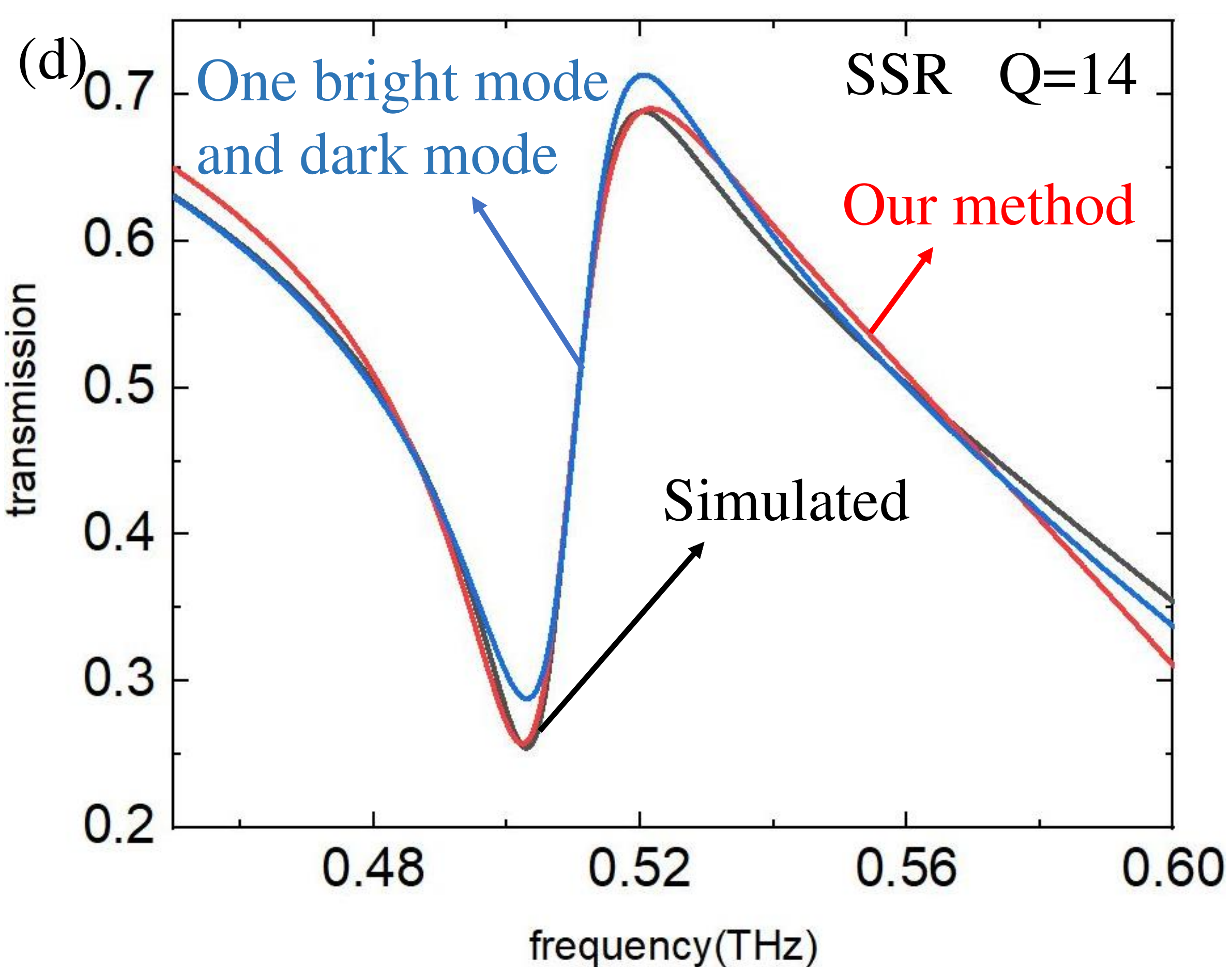}
		\includegraphics[width=0.315\textwidth]{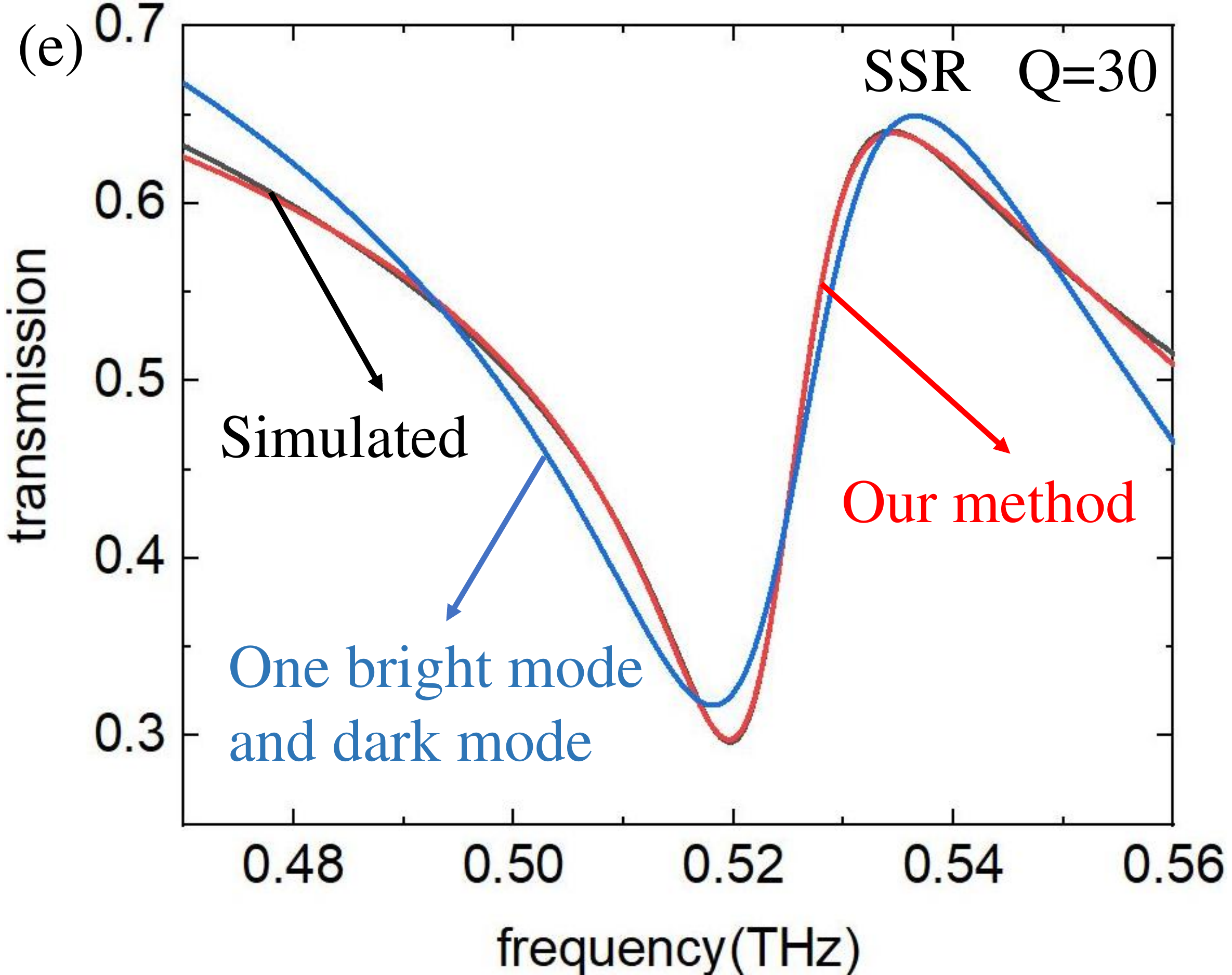}
		\includegraphics[width=0.315\textwidth]{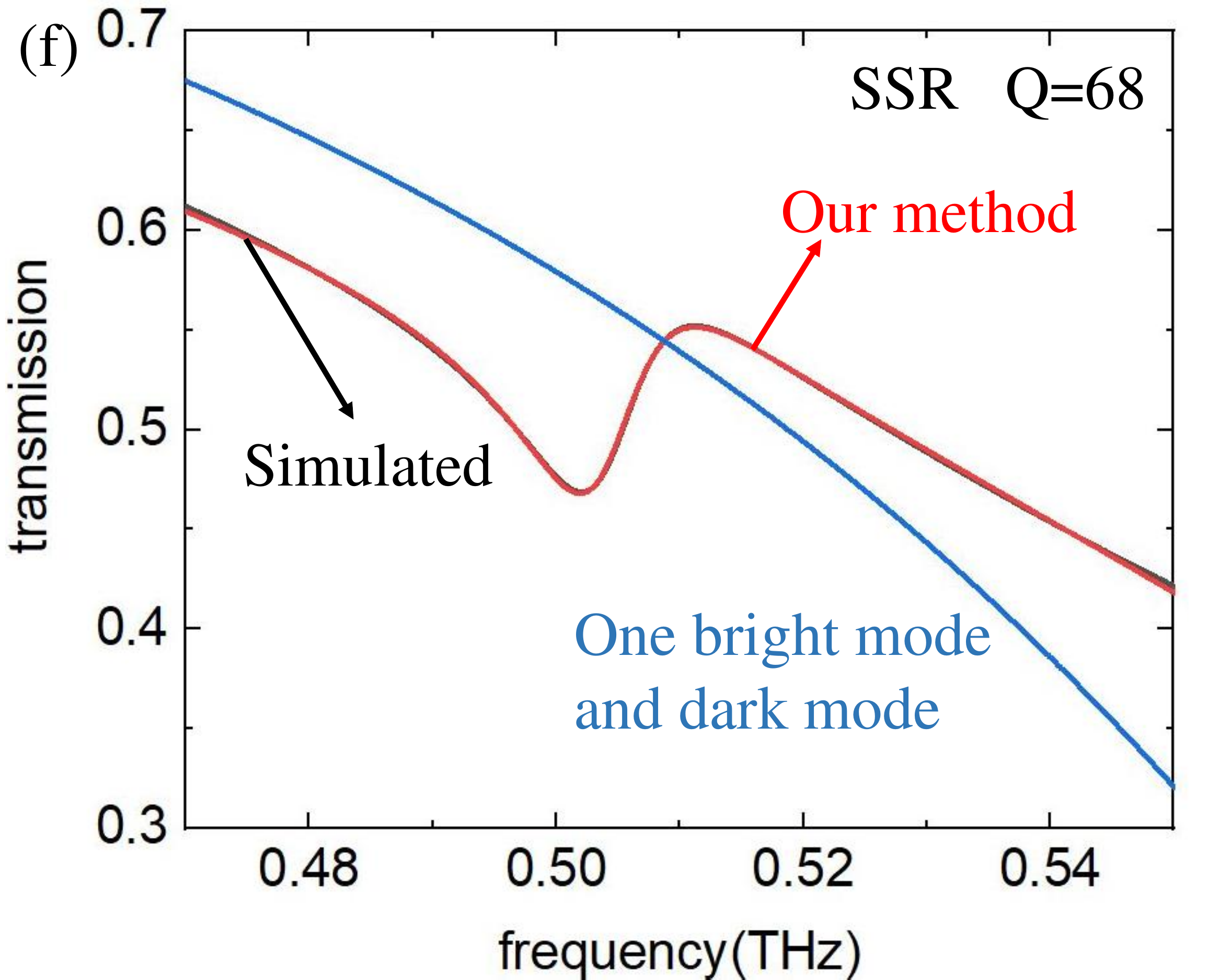}
	\caption{The black lines, blue lines and red lines demonstrate the full-wave simulations, the fitting results of previous theory \cite{Cong2015} and  the fitting results of our theory, respectively. The (a), (b) and (c) represent the low-Q, medium Q and high Q cases for CW BIC (see Fig. \ref{Fig1} (a)). The (d), (e) and (f) show the low-Q, medium Q and high Q cases for SRR BIC (see Fig. \ref{Fig1} (b)).}
	\label{Fig6}
\end{figure*}

\begin{equation}
\chi_{\text{eff}} =\frac{ \sqrt{\gamma_{a}}a+\sqrt{\gamma_{b}}e^{_{i\phi}}b} {\epsilon_{0} E }.
\end{equation}
Finally, we obtain the transmission spectrum with $T \approx 1- \text{Im}(\chi_{\text{eff}})$ \cite{Cong2015} , as shown,
\begin{widetext}
\begin{equation}
	T \approx 1-\text{Im}(\frac{ (w-w _{a}-i\gamma_{a}) \gamma_{b}e^{_{2i\phi}}+((w-w _{b}-i\gamma_{b}) \gamma_{a}-2\Omega \sqrt{\gamma_{a}\gamma_{b}}e^{_{i\phi}}} { ( w-w _{b}-i\gamma_{b}) (w-w _{a}-i\gamma_{a})-\Omega^{2}}).
\end{equation}
\end{widetext}

\section{Examples}
Firstly, we present the BIC with coupling between two CWs (as shown in Fig. \ref{Fig1} (a)) and we can obtain the resonance frequency ($\omega_1$, $\omega_2$), the loss ($\gamma_1$, $\gamma_2$) and phase with each resonance frequency ($\phi_1$, $\phi_2$) for each single metamaterial structure, as shown in Fig. \ref{Fig2}. We use the full-wave simulations of each metamaterial structure and put them together for BIC to get the corresponding transmission spectrum (see the black line of Fig. \ref{Fig3}). 

From the results, we can obtain the resonance frequency  $\omega_1 = 0.256$ THz, $\omega_2 = 0.248$ THz, the loss $\gamma_1 = 0.028$ THz, $\gamma_2 = 0.03$ THz and phase with each resonance frequency $\phi_1 =0.297 $, $\phi_2 =0.31 $ for each single metamaterial structure. Therefore, $\omega_a = 0.256$ THz, $\omega_b = 0.248$ THz, $\gamma_a = 0.028$ THz, $\gamma_b = -0.004$ THz and $\phi =2.3 $. From the result of BIC (as shown in black line of Fig. \ref{Fig3}), we observe the BIC phenomenon with Fano resonance transmission spectrum with $Q=w_{BIC}/ \Delta w = 160$, where $\Delta w$  is the full width at half maximum (FWHM) of Fano resonance. 

Due to complex analytic calculation of coupling strength $g$ between two metamaterial structures, we employ the fitting calculation by applying our universal coupled theory (Eq. 4) to obtain the coupling strength $g$, as shown in the red line of Fig. \ref{Fig3}. As we can see that the transmission spectrum of our theory (Eq. 4) is consistent with the Full-wave simulation of CWs BIC. Therefore, our theory can not only well describe and explain BIC for metamaterial but also can predict the frequency and Q-value of BIC. 

In order to demonstrate the universality of our theory for different coupled metamaterial structures, we verify our theory by using coupling between two SRR as BIC as shown in Fig. \ref{Fig1} (b). We take the same steps above where get the Full-wave simulations of each single SRR (see Fig. \ref{Fig4}) and BIC (see Fig. \ref{Fig1} (b)) to obtain the corresponding transmission spectrum (the spectrum of each single structure see Fig. \ref{Fig4} and the spectrum of BIC shows black line of Fig. \ref{Fig5}). As we can see from Fig. \ref{Fig4}, $\omega_a = \omega_1 = 0.573$ THz, $\omega_b =\omega_2 = 0.636$ THz, $\gamma_a = \gamma_1 = 0.144$ THz, $\gamma_b = (\gamma_{1} - \gamma_{2})/2 = -0.008$ THz, $\phi = (\phi_{1} - \phi_{2})d =2.1 $ and the transmission spectrum of two SRRs BIC with $Q = 18$. Subsequently, we employ our universal coupled theory (Eq. 4) with those parameters by fitting coupling strength $g$ to calculate the transmission spectrum for BIC, as shown in red line in Fig. \ref{Fig5}. It is easy to find that our method is very closely to Full-wave simulations of two SRRs BIC. In other words, our universal coupled theory is valid for BIC with different types of metamaterial structures.

\section{Discussion}

Our method proposes the two bright modes coupling and employs it into the BIC metamaterial. Therefore, our theory can easily predict the transmission spectrum of BIC and much more suitable for high Q resonance cases comparing with previous research \cite{Cong2015}. In order to demonstrate the superiority of our theory, we employ full-wave simulations of low Q, medium Q and high Q cases for CW BIC and SRR BIC, respectively, as shown in the black lines of Fig. \ref{Fig6}. Subsequently, we apply the coupling of one bright mode and one dark mode \cite{Cong2015} and our theory (the coupling of two bright modes) to fit the full-wave simulations of low Q, medium Q and high Q cases respectively. The blue lines demonstrate the fitting results of the previous theory \cite{Cong2015} and the red lines show the fitting results of our theory in Fig. \ref{Fig6}. 

From the results of Fig. \ref{Fig6}, we can easily obtain that the fitting results of the previous theory and our theory can both well predict the low Q case of the transmission spectrum of BIC, as shown in Fig. \ref{Fig6} (a) and (d). When the Q-value continuously increases, the fitting results of the previous theory start to have some errors in the medium Q cases, as shown in the blue lines in Fig. \ref{Fig6} (b) and (e). When the high Q-value cases for BIC (see Fig. \ref{Fig6} (c) and (f)) occurs, the previous theory of the coupling of one bright mode and one dark mode can not be valid anymore and the previous theory can not give the Fano spectrum of BIC at all. The reason for this phenomenon is that when the two metamaterial structures become increasingly uniform ($\Delta l$ turns smaller and smaller), the Q-value of BIC becomes larger and larger. When the high Q cases occur, two metamaterial structures can be excited by the input THz wave. Thus, the two metamaterial structures turn into two bright modes and the theory of one bright mode and dark mode can not be valid anymore. However, the fitting results of our theory have good performance of fitting results to predict BIC for low Q, medium Q and high Q cases, as shown in red lines of Fig. \ref{Fig6}. Therefore, our theory surpasses and contains the previous theory \cite{Cong2015}, especially for high Q-value resonance BIC. 

\section{Conclusion}
In this paper, we propose a new theory to describe metamaterial BIC by employing the coupling theory of two bright modes with phase. Our new theory can universally explain metamaterial BIC and very accurately predict the frequency of BIC and Q-value of resonance, even for the very high Q case.

\section*{Acknowledgements}
This work acknowledges funding from National
Key Research and Development Program of China
(2019YFB2203901); National Science and Technology
Major Project (grant no: 2017ZX02101007-003); National
Natural Science Foundation of China (grant no: 61565004;
61965005; 61975038; 62005059). The Science and
Technology Program of Guangxi Province (grant no:
2018AD19058). W.H. acknowledges funding from Guangxi
oversea 100 talent project; W.Z. acknowledges funding from
Guangxi distinguished expert project.


\begin{thebibliography}{99}
\bibitem{Neumann1929}  J. von Neumann and E. Wigner, “Uber merkw¨urdige diskrete eigenwerte,” Z. Physik 50, 291-293 (1929).

\bibitem{Hsu2016} C. W. Hsu, B. Zhen, A. D. Stone, J. D. Joannopoulos, and M. Soljaˇci´c, “Bound states in the continuum,” Nat. Rev. Mater. 1, 16048 (2016).

\bibitem{Gao2016} X. Gao, C. W. Hsu, B. Zhen, et al., “Formation mechanism of guided resonances and bound states in the continuum in photonic crystal slabs,” Sci. Rep., vol. 6, p. 31908, 2016.

\bibitem{Gansch2016}  R. Gansch, S. Kalchmair, P. Genevet, et al., “Measurement of
bound states in the continuum by a detector embedded in a
photonic crystal,” Light Sci. Appl., vol. 5, no. 9, p. e16147,

\bibitem{Yang2016}  Y. Yang, C. Peng, Y. Liang, Z. Li, and S. Noda, “Analytical
perspective for bound states in the continuum in photonic crystal
slabs,” Phys. Rev. Lett., vol. 113, no. 3, 2014, Art no. 037401.
2016.

\bibitem{Bulgakov2008} E. N. Bulgakov and A. F. Sadreev, “Bound states in the continuum
in photonic waveguides inspired by defects,” Phys. Rev. B, vol.
78, 2008, Art no. 075105.

\bibitem{Paddon2000}  P. Paddon and J. F. Young, “Two-dimensional vector-coupled-mode theory for textured planar waveguides,” Phys. Rev. B, vol.
61, pp. 2090–2101, 2000.

\bibitem{Pacradouni2000}  V. Pacradouni, W. J. Mandeville, A. R. Crown, P. Paddon,
J. F. Young, and S. R. Johnson, “Photonic band structure of
dielectric membranes periodically textured in two dimensions,”
Phys. Rev. B, vol. 62, p. 4204, 2000.

\bibitem{Azzam2018}  S. I. Azzam, V. M. Shalaev, A. Boltasseva, and A. V. Kildishev,
“Formation of bound states in the continuum in hybrid
plasmonic-photonic systems,” Phys. Rev. Lett., vol. 121, no. 25,
p. 253901, 2018.

\bibitem{Yu2019} Yu, Zejie, et al. "Photonic integrated circuits with bound states in the continuum." Optica 6.10 (2019): 1342-1348.

\bibitem{Chen2019} Chen, Junxue, and Peixin Chu. "Phase-induced Fano antiresonance in a planar waveguide with two dielectric ridges." JOSA B 36.12 (2019): 3417-3427.

\bibitem{Lee2020} Lee, Sun-Goo, Seong-Han Kim, and Chul-Sik Kee. "Bound states in the continuum (BIC) accompanied by avoided crossings in leaky-mode photonic lattices." Nanophotonics 1.ahead-of-print (2020).

\bibitem{Bykov2019} Bykov, Dmitry A., Evgeni A. Bezus, and Leonid L. Doskolovich. "Coupled-wave formalism for bound states in the continuum in guided-mode resonant gratings." Physical Review A 99.6 (2019): 063805.

\bibitem{Gao2019} X. Gao, B. Zhen, M. Soljačić, H. Chen, and C. W. Hsu, “Bound
states in the continuum in fiber Bragg gratings,” ACS Photonics,
vol. 6, pp. 2996–3002, 2019.

\bibitem{Koshelev2018} Koshelev, Kirill, et al. "Asymmetric metasurfaces with high-Q resonances governed by bound states in the continuum." Physical review letters 121.19 (2018): 193903.

\bibitem{Miroshnichenko2010} Miroshnichenko, Andrey E., Sergej Flach, and Yuri S. Kivshar. "Fano resonances in nanoscale structures." Reviews of Modern Physics 82.3 (2010): 2257.

\bibitem{Kupriianov2019} A. S. Kupriianov, Y. Xu, A. Sayanskiy, V. Dmitriev, Y. S. Kivshar,
and V. R. Tuz, “Metasurface engineering through bound states in
the continuum,” Phys. Rev. Appl., vol. 12, no. 1, 2019, Art no.
014024.

\bibitem{Abujetas2019} D. R. Abujetas, N. van Hoof, S. ter Huurne, J. G. Rivas, and
J. A. Sánchez-Gil, “Spectral and temporal evidence of robust
photonic bound states in the continuum on terahertz
metasurfaces,” Optica, vol. 6, no. 8, pp. 996–1001, 2019.

\bibitem{Cong2015} Cong, Longqing, et al. "Fano resonances in terahertz metasurfaces: a figure of merit optimization." Advanced Optical Materials 3.11 (2015): 1537-1543.

\bibitem{Huang2014} Huang, Wei, Andon A. Rangelov, and Elica Kyoseva. "Complete achromatic optical switching between two waveguides with a sign flip of the phase mismatch." Physical Review A 90.5 (2014): 053837.

\bibitem{Huang2019} Huang, Wei, et al. "Robust and broadband integrated terahertz coupler conducted with adiabatic following." New Journal of Physics 21.11 (2019): 113004.

\bibitem{Huang2020} Huang, Wei, et al. "Quantum Engineering Enables Broadband and Robust Terahertz Surface Plasmon-Polaritons Coupler." IEEE Journal of Selected Topics in Quantum Electronics 27.2 (2020): 1-7.

\bibitem{Huang20202} Huang, Wei, et al. "Long-distance adiabatic wireless energy transfer via multiple coils coupling." Results in Physics 19 (2020): 103478.


\bibitem{Liang2020} Liang, Yao, et al. "Bound states in the continuum in anisotropic plasmonic metasurfaces." Nano Letters 20.9 (2020): 6351-6356.

\bibitem{Kodigala2017}  A. Kodigala, T. Lepetit, Q. Gu, B. Bahari, Y. Fainman, and
B. Kanté, “Lasing action from photonic bound states in
continuum,” Nature, vol. 541, pp. 196–199, 2017.

\bibitem{Ha2018} S. T. Ha, Y. H. Fu, N. K. Emani, et al., “Directional lasing in
resonant semiconductor nanoantenna arrays,” Nat.
Nanotechnol., vol. 13, pp. 1042–1047, 2018.

\bibitem{Liu2017} Y. Liu, W. Zhou, and Y. Sun, “Optical refractive index sensing
based on high-Q bound states in the continuum in free-space
coupled photonic crystal slabs,” Sensors, vol. 17, no. 8, p. 1861,
2017.

\bibitem{Romano2018} S. Romano, A. Lamberti, M. Masullo, et al., “Optical biosensors
based on photonic crystals supporting bound states in the
continuum,” Materials, vol. 11, no. 4, p. 526, 2018.

\bibitem{Khanikaev2013} Khanikaev, Alexander B., Chihhui Wu, and Gennady Shvets. "Fano-resonant metamaterials and their applications." Nanophotonics 2.4 (2013): 247-264.

\bibitem{Tittl2018} Tittl, Andreas, et al. "Imaging-based molecular barcoding with pixelated dielectric metasurfaces." Science 360.6393 (2018): 1105-1109.

\bibitem{Foley2014} J. M. Foley, S. M. Young, and J. D. Phillips, “Symmetry-protected
mode coupling near normal incidence for narrow-band
transmission filtering in a dielectric grating,” Phys. Rev. B, vol.
89, no. 16, p. 165111, 2014.


\bibitem{Zhen2015} Zhen, Bo, et al. "Spawning rings of exceptional points out of Dirac cones." Nature 525.7569 (2015): 354-358.

\bibitem{Zhen2014} Zhen, Bo, et al. "Topological nature of optical bound states in the continuum." Physical review letters 113.25 (2014): 257401.

\bibitem{Limonov2017} Limonov, Mikhail F., et al. "Fano resonances in photonics." Nature Photonics 11.9 (2017): 543.

\bibitem{Gallinet2011} Gallinet, Benjamin, and Olivier JF Martin. "Ab initio theory of Fano resonances in plasmonic nanostructures and metamaterials." Physical Review B 83.23 (2011): 235427.

\bibitem{zhang2014} Zhang F, Huang XC, Zhao Q,et al. "Fano resonance of an asymmetric dielectric wire pair." Applied Physics Letters, 2014, 105(17):172901.

\bibitem{meng2012} Meng F Y, Wu Q, Erni D, et al. "Polarization-Independent Metamaterial Analog of Electromagnetically Induced Transparency for a Refractive-Index-Based Sensor." IEEE Transactions on Microwave Theory and Techniques, 2012, 60(10):3013-3022.

\end{thebibliography}
\end{document}